\documentclass[useAMS,usenatbib]{mn2e} 
\usepackage{enumitem}
\usepackage{graphicx}
\usepackage{amsmath}
\usepackage{amssymb}
\usepackage{color}
\usepackage{natbib}
\usepackage{threeparttable}
\usepackage{changes}
\usepackage{aas_macros}
\def\actaa{Acta. Astron.}
\usepackage{url}
\usepackage{fancyref}
\usepackage[colorlinks=true,linkcolor=blue,citecolor=blue]{hyperref}
 
\title[X-ray and UV emission from populations of accreting white
  dwarfs in elliptical galaxies]{Population synthesis of accreting 
  white dwarfs: II. X-ray and UV emission}
\author[Hai-Liang Chen, T. E. Woods, L. R. Yungelson, M. Gilfanov and Zhanwen
Han]{Hai-Liang Chen$^{1,2,3,4}$\thanks{E-mail:
chenhl@ynao.ac.cn}, T. E. Woods$^{2}$, L. R. Yungelson$^{5}$, M.
Gilfanov$^{2,6,7}$, Zhanwen Han$^{1,3}$\\
  $^{1}$Yunnan Observatories, Chinese Academy of Sciences, Kunming, 650011,
China\\
  $^{2}$Max Planck Institute for Astrophysics, Karl-Schwarzschild-Str. 1,
Garching b. M{\" u}nchen 85741, Germany\\
  $^{3}$Key Laboratory for the Structure and Evolution of Celestial Objects, 
Chinese Academy of Sciences, Kunming 650011, China\\
  $^{4}$University of Chinese Academy of Sciences, Beijing 100049, China\\
  $^{5}$Institute of astronomy, RAS, 48 Pyatnitskaya Str., 119017 Moscow, Russia\\
  $^{6}$Space Research Institute of Russian Academy of Sciences, Profsoyuznaya
84/32,117997 Moscow, Russia\\
  $^{7}$Kazan Federal University, Kremlevskaya str.18, 420008 Kazan, Russia\\  
}
\begin{document}


\pagerange{\pageref{firstpage}--\pageref{lastpage}} \pubyear{2013}

\maketitle

\label{firstpage}

\begin{abstract}

Accreting white dwarfs (WDs) with non-degenerate companions are expected to
emit in soft X-rays and the UV, if accreted H-rich material burns stably.
They are an important component of the unresolved emission of
elliptical galaxies, and their combined ionizing luminosity may
significantly influence the optical line emission from warm ISM. In
an earlier paper we modeled populations of accreting WDs, first generating 
WD with main-sequence, Hertzsprung gap and red giant companions 
with the population synthesis code \textsc{BSE}, and then following 
their evolution with a grid of evolutionary tracks computed with \textsc{MESA}.
Now we use these results to estimate the soft X-ray
(0.3-0.7keV), H- and He II-ionizing luminosities of nuclear burning
WDs and the number of super-soft X-ray sources for galaxies
with different star formation histories. For the starburst case, these
quantities peak at $\sim 1$ Gyr and decline by $\sim 1-3$ orders of
magnitude by the age of 10 Gyr.  For stellar ages of $\sim$~10 Gyr,
predictions of our model are consistent with soft X-ray luminosities
observed by Chandra in nearby elliptical galaxies and He II
4686$\AA/\rm{H}{\beta}$ line ratio measured in stacked SDSS spectra of
retired galaxies, the latter characterising the strength and hardness
of the UV radiation field. However, the soft X-ray luminosity and
He~II~4686$\AA/\rm{H}{\beta}$ ratio are significantly overpredicted  
for stellar ages of $\lesssim 4-8$ Gyr. 
We discuss various possibilities to resolve this discrepancy and tentatively
conclude that it may be resolved by a modification of   
the typically used criteria of dynamically unstable mass loss for 
giant stars.

\end{abstract}



\section{Introduction}
\label{sec:intro}

Accreting white dwarfs (WDs) are important for the study of binary
evolution and accretion physics (see \citealt{py14,mmn14} for a
recent review). Depending on the accretion rates, they are deemed to
be components of cataclysmic binaries \citep[see][for a comprehensive
  review]{warn03} and supersoft X-ray sources \citep{vbnr92}.  Under
differing initial conditions, they are the progenitors of 
components of future double white dwarfs, type Ia supernovae and binary millisecond pulsars
\citep[e.g.][]{wi73,ty81,webb84,it84,tv86}.  Although they are widely
studied, there remain many uncertainties 
concerning their formation and
evolution. One avenue to constrain this problem is to invoke binary
population synthesis. This allows one to compare the observed numbers
and characteristics of these objects with that which are expected
given our present understanding of binary evolution and accretion
physics.

Hydrogen-rich material accreted by a WD from its companion may burn at
its surface either stably or unstably. The result is strongly
dependent on the mass transfer rate 
and mass of the WD \citep{pz78,sat79}.  If the accretion
rate corresponds to the stable nuclear 
burning regime, accreting white dwarfs (henceforth, SNBWD) have
typical effective temperatures of $10^{5} - 10^{6}$K, emitting
prominently in the soft X-ray and EUV.  In this regime, not all
accreting WDs can be observed as SSSs, as this depends on their
intrinsic luminosity, effective temperature and the column density of
the gas along the line of sight. In the present paper, we refer to
SNBWDs with observed soft X-ray (0.30-0.70~keV) luminosity $L_{\rm x}
> 10^{36}$~ erg/s as SSSs.  For accretion rates below the stability
limit, hydrogen burning occurs only in unstable flashes, observable as
novae. There will be little X-ray emission in this regime, except in
the post-novae SSS phase (\citet{tg95,wbbp13}, Soraisam et al. (in prep.)).
If the accretion rate is
larger than the maximum rate for stable nuclear-burning, the result is
still unclear \citep{cit98,hkn96}.  In the scenario of \citet{cit98},
the WD will become a red giant after accreting a small amount of mass
\citep{pacz71}.  On the other hand, \citet{hkn96} proposed that in
this regime the excess mass is lost as an optically thick
wind. Following \citet{lv13}, the WDs in this regime are dubbed
rapidly accreting white dwarfs (RAWDs).  In this scenario, an
accreting WD's photosphere will expand modestly, lowering their
effective temperatures such that they radiate predominantly in the
EUV.

In addition to the emission from X-ray binaries dominating the X-ray
radiation in most galaxies, extended emission in soft X-ray band
is widely observed in
galaxies of different morphologies. However, the origin of the
extended emission is still unclear. \citet{bg08} demonstrated that the
emission from accreting WDs should be an important component of the
unresolved emission.

A population of accreting WDs with effective temperatures $T_{\rm eff}
> 10^{5}$K should be capable of ionizing the interstellar medium.
\citet{rckm94} were the first to model ionization nebulae around SSSs,
for the simple case of spherical symmetry and blackbody spectra.  They
found that these nebulae should have strong He II $\lambda$4686, [O
  III] $\lambda 5007$, and [O I] $\lambda 6300$ emission
lines. \citet{wg13} demonstrated that accreting WDs may provide a
significant, or even dominant, contribution to the ionizing
background, particularly in early-type galaxies with
passively-evolving stellar populations (those with negligible ongoing
star formation). Depending on their numbers, this could significantly
impact the ionization state of the warm ISM in gas-rich early-type
galaxies.  \citet{soma+12} found that many early type galaxies (at
least 40\% outside the Virgo cluster) host a detectable mass of HI,
i.e. $\sim(10^{8}-10^{9}) M_{\odot}$.  Morphologically, this gas is
found in regular HI discs, rings and irregularly distributed
clouds. Most common are HI discs and rings, which may extend to tens
of kpc, suggesting a covering factor of $\sim$1/2 for the ionizing
radiation from the stellar population.

Although emission from populations of accreting WDs is important, it
has been seldom studied.  In Paper I \citep{cwyg+14}, we conducted a
population synthesis study of accreting white dwarfs. We found that it
is important to have a realistic prescription for mass transfer and
investigated the number of RAWDs, SNBWDs and the SNe Ia rate in the
single degenerate (SD) scenario.  In this paper, based on the results
of our binary population synthesis of Paper I, we study the soft X-ray
and UV emission of the population of accreting WDs and their contribution
to the radiation of elliptical galaxies in the He II 4686$\AA$ and
H$\beta$ emission lines. These results are then compared with
observations.

The paper is structured as follows. In section \ref{sec:emi}, we
describe how we calculate the emission spectra of accreting WDs.  In
section \ref{sec:bps}, we briefly outline the typical 
assumptions underlying our population synthesis.  In section
\ref{sec:res}, we present the soft X-ray (0.3-0.7keV) luminosity, H
ionizing luminosity, and He II ionizing luminosity from the population
of accreting WDs as a whole. We also compute the expected ratio
He~II4686$\AA$/H$\beta$ as a function of age for a starburst stellar
population and compare with observations of early type galaxies in the
Sloan Digital Sky Survey \citep[SDSS;][]{jwgs+14}. In section
\ref{sec:disc}, we discuss the uncertainties in our standard model and
present a potential solution to resolve the discrepancy between the
results found in our standard model and observations. In addition, we show
the evolution of SSS number as a function of stellar age. Finally, we
summarize our results and conclude in section \ref{sec:con}.


\section{Emission spectra of accreting white dwarfs}
\label{sec:emi}

\begin{figure}               
\centering
\includegraphics[width=84mm]{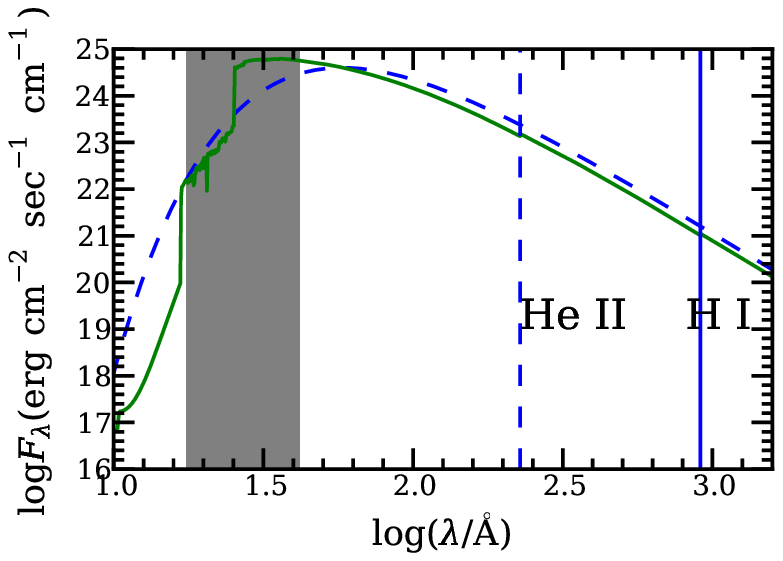}
\caption{Comparison of an accreting WD spectrum computed by means of a
  NLTE model (green solid line, \citet{wern86,wern89,rauch03}) with
  that found from the blackbody approximation (blue dashed line) for a
  WD with effective temperature $T_{\rm eff}=5.0\times10^{5}$ K,
  log$(\rm{g/cm/s^{2}}) = 8.0$.  The two vertical lines represent the
  H~I and He~II photoionizing limits. The edges around $25$ \AA\, and
  $15$ \AA\, are the C VI (25.30\AA) and O VII (16.77\AA) absorption
  edges, respectively. The shaded area denotes the soft X-ray band
  (0.3-0.7keV) in our calculation.  The NLTE spectrum was retrieved
  from TheoSSA (http://dc.g-vo.org/theossa).}
\label{fig:wd_spect}
\end{figure}

Detailed synthetic spectral models for accreting WDs can be produced
using a non-local thermodynamic equilibrium (NLTE) stellar atmosphere
model \citep[e.g.][]{wern86,wern89,rauch03}, which is known to
approximately reproduce SSSs spectra \citep[e.g.][]{rogn+10}. In
Fig. \ref{fig:wd_spect}, we compare the emission spectra of a SNBWD
computed by means of a NLTE model\footnote{http://dc.g-vo.org/theossa}
with the blackbody approximation. Except for some absorption lines,
the difference is not dramatic, particularly near the H and He II
absorption edges. The difference in soft X-ray (0.3-0.7)\,keV luminosity
between the blackbody approximation and NLTE models is smaller for
accreting WDs with high temperatures and greater for low
temperatures. With this discrepancy in mind, we adopt the blackbody
approximation for SNBWD spectra in this paper.

For a SNBWD, the nuclear burning luminosity is
\begin{equation}
L_{\rm nuc} = {\epsilon_{\rm H}} {X_{\rm H}} {\dot{M}_{\rm acc}},
\end{equation}
where $\epsilon_{\rm H} = 6.4\times10^{18}$ erg/g is the nuclear
energy release per unit mass of hydrogen, $X_{\rm H}$ is the mass fraction
of hydrogen and $\dot{M}_{\rm acc}$ is the accretion rate. We ignore
the accretion luminosity in the stable-burning regime, since it is
much lower (less than 10\% for $M_{\rm WD} < 1.2M_{\odot}$ and 20\%
for $1.20M_{\odot} < M_{\rm WD} < 1.4M_{\odot}$) compared with the
nuclear burning luminosity.

Detailed models of SNBWDs show that their photospheric radii are about
$1-8$ times as large as the radius of a cool WD \citep[depending on
  the accretion rates,][]{iben82,wbbp13}. \cite{wbbp13} simulated the
evolution of accreting WDs with different accretion rates using the
\textsc{MESA} code \citep{pbdh+11,pcab+13}. From their results, we
know the dependence of the photospheric radius on the accretion rates
in the stable-burning regime.  For any given WD mass and accretion
rate, we find the radius by linear interpolation of the results of
\citet{wbbp13} between WD mass and accretion rate. Then the effective
temperature of SNBWDs is found from the Stefan-Boltzmann law.

 RAWDs with accretion rates larger than the maximum stable burning
rate, expand as the WD loses mass and the effective
temperature decreases. \cite{hkn99} investigated the optically-thick
wind solution for different white dwarfs with different accretion
rates. They showed the dependence of photospheric radius and effective
temperature on the accretion rate (see their Figs. 3, 4).  By linear
fitting of their results and linear interpolation among different WD
masses, we can find the radius and the effective temperature for any
combination of white dwarf mass and accretion rate. It has been shown
that RAWDs with typical mass accretion rates (less than
$4.0\times10^{-6}M_{\odot}\rm{yr}^{-1}$) and effective temperatures 
(T $\approx$ 1--2$\times 10^{5}$K) can fully ionize hydrogen and helium beyond the wind photosphere
\citep{wg13}. For high accretion rates, the spectra of RAWDs are not
well approximated by blackbody spectra because of absorption.
However, high accretion rates 
 can only be sustained for a short time and will
contribute little to the total luminosity. Therefore, it is reasonable
to use blackbody spectra as an approximation for RAWDs.


\section{Binary Population Synthesis}
\label{sec:bps}

In Paper I, we have introduced our approach to studying populations of
accreting WDs, and have shown the importance of a careful treatment of
the second mass transfer phase.  Following the method outlined in
Paper I, we carry out two sets of calculations with our
\textsc{BSE}+\textsc{MESA} model. Here we briefly summarize the
procedure and main assumptions of the calculation.

First, we evolve a set of binaries with different initial parameters
using the \textsc{BSE} code \citep{hpt00,htp02}, and obtain the binary
parameters at the beginning of mass transfer for a population of WD
binaries with main-sequence, Hertzsprung gap and red giant
companions. Second, we compute a WD binary evolution library by means
of the detailed stellar evolution code \textsc{MESA}
\citep{pbdh+11,pcab+13}, which describes the evolution of WD binaries
with different parameters at the beginning of mass transfer. For each
WD binary with a given WD mass, donor mass and orbital period from the
set of systems obtained in the first step of computations, we select
the nearest track in the WD binary evolution library in order to
follow its subsequent evolution. We take 1.2 $M_{\odot}$ as the upper
limit of the nascent CO WD, which was implemented in the \textsc{BSE}
code, although modern estimates give $\approx 1.07 M_{\odot}$ for
single stars with Z = 0.02 \citep[e.g.][]{dgsl+15}.  For the
distribution of initial binary parameters, we use the Kroupa IMF for
the primary mass \citep{krou01}, a flat mass ratio distribution
\citep{kpty79} and a flat distribution in logarithmic space for binary
separation \citep{abt83}.  To describe common envelope evolution, we
employ the $\alpha$-formalism \citep{webb84,deko90}.  We use the
fitting formula for the value of the binding energy parameter
$\lambda$ from \citet{lvk11}.

The efficiency $\alpha$ of the common envelope phase in ejecting the
donor star's envelope remains uncertain \citep[see][for a
  comprehensive review]{ijcd+13}.  The latest studies of post-common
envelope binaries \citep[e.g.][]{dkk12} restrict $\alpha$ to a
canonical value $\le 1$ with a tendency to decrease with increasing
secondary mass and WD mass. Some studies
(e.g. \citet{dkw10,zsgn10,rt12}) show that, likely, $\alpha <
0.50$. Here we adopt two values: $\alpha = 0.25$ (model a025) and
$\alpha = 0.50$ (model a050), performing calculations for each. In
Paper I, we verified that the population of hydrogen-accreting WDs
depends relatively weakly on the assumed value of $\alpha$.

Combining our binary population synthesis results for accreting white
dwarfs with their predicted emission spectra, we study their X-ray and
UV emission. As in Paper I, we investigate two simple but
representative cases. (I) Starburst: all stars are formed at $t = 0$,
after which the star formation rate is zero. (II) Constant star
formation rate for 10 Gyr. These two cases provide us with simple
models which are representative of elliptical-like and spiral-like
galaxies, respectively.


\section{Results}
\label{sec:res}

\subsection{X-ray emission of accreting white dwarfs}
\label{sec:xray}

\begin{figure}   
\centering 
\includegraphics[width=84mm]{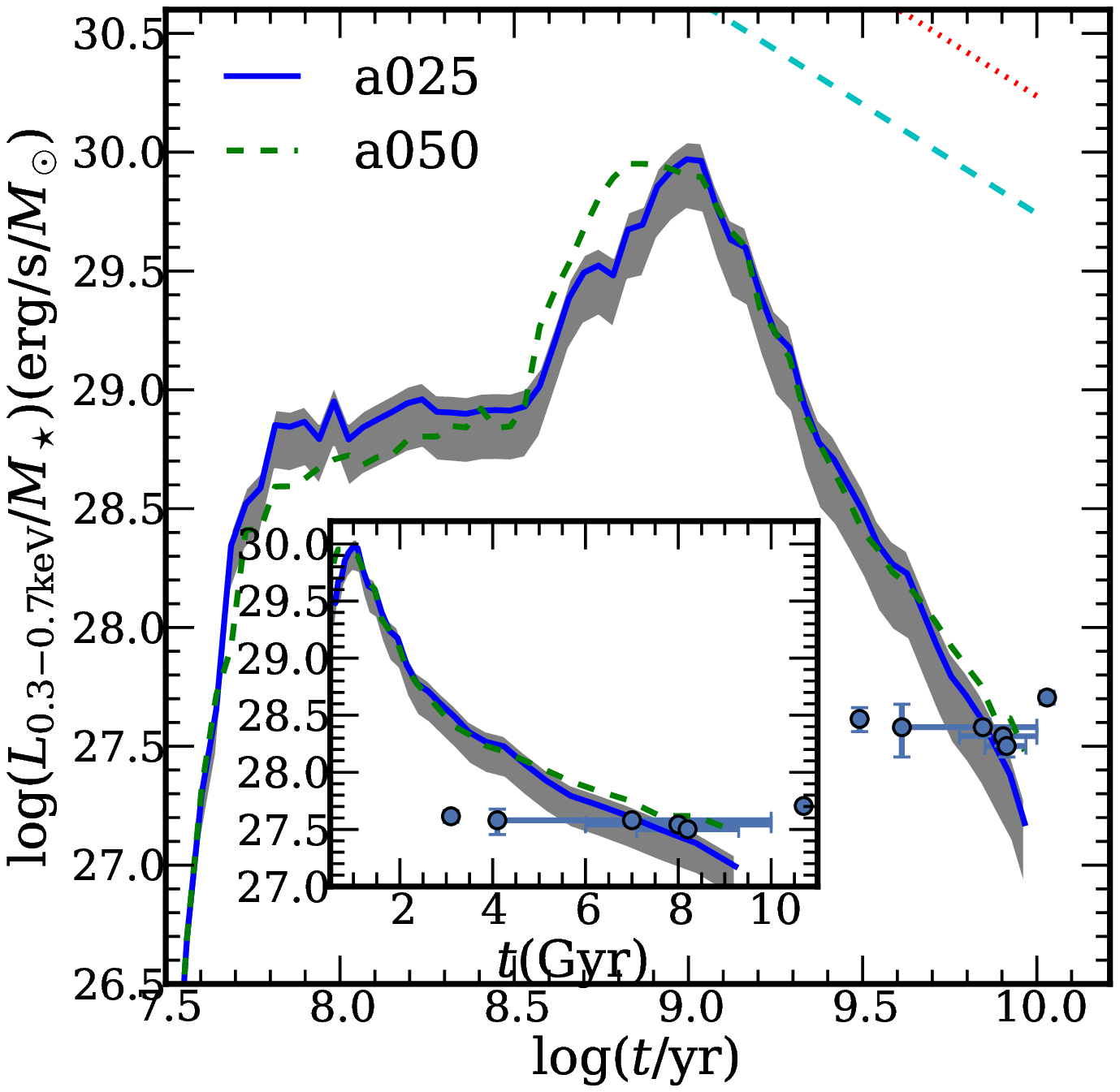}
\includegraphics[width=84mm]{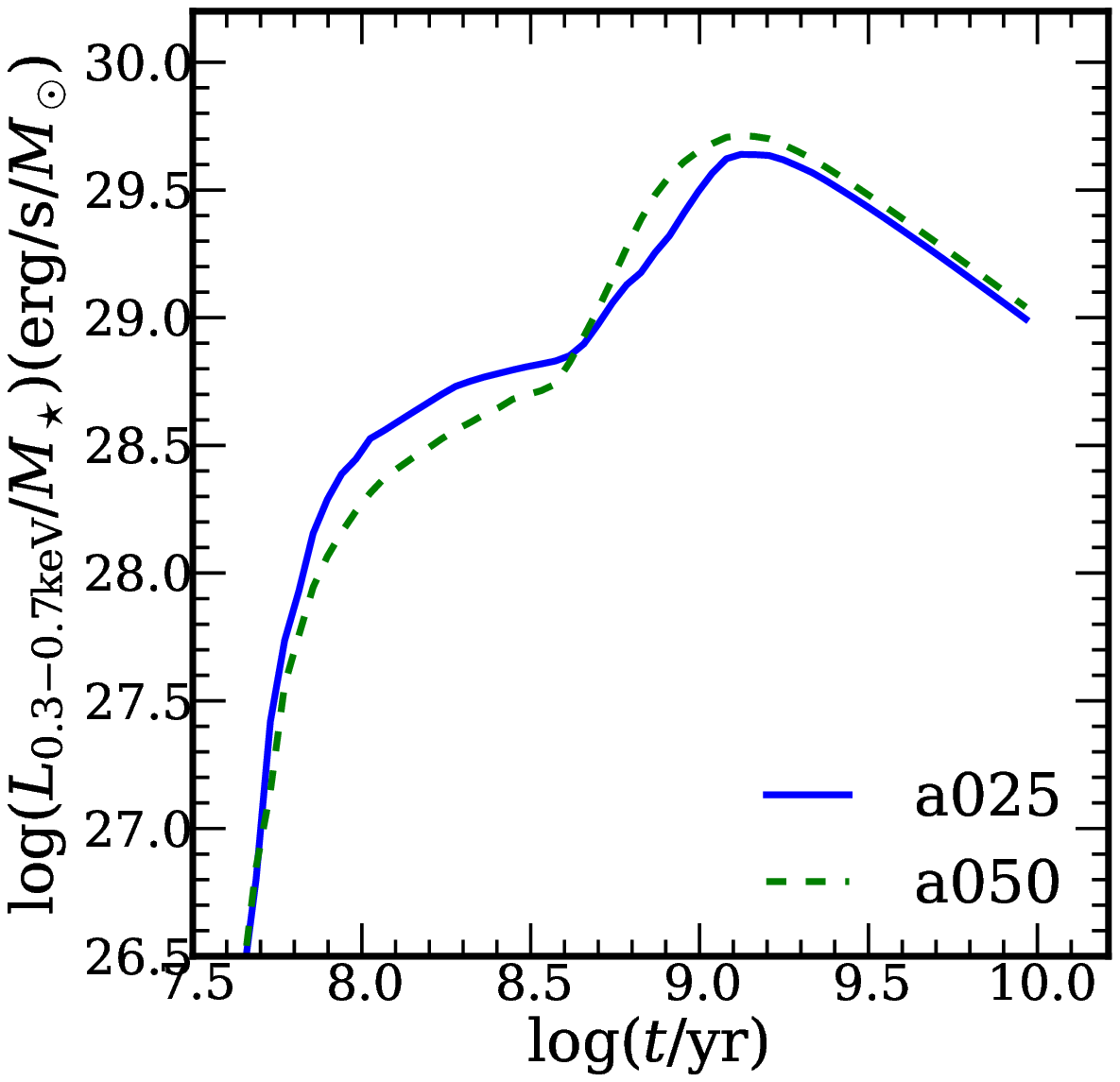}
\caption{
Mass-normalized X-ray luminosity (absorption applied with
  $N_{\rm H}=3.0\times10^{20} {\rm cm}^{-2}$) in soft (0.3-0.7keV)
  band for starburst case (upper panel) and constant SFR case (lower
  panel) as a function of stellar age. In the upper panel, the shaded area shows the X-ray
  luminosity with $1.8\times10^{20}\rm{cm}^{-2}<N_{\rm
    H}<6.7\times10^{20}\rm{cm}^{-2}$. 
The blue solid and green dashed lines are for models a025 and a050, respectively.
The points with error bars are
  the observed X-ray luminosities for individual elliptical galaxies 
  \citep[see Table \ref{tab:nsl},][]{bg10,zgb12}. 
 The cyan dashed line is
  the expected soft X-ray luminosity if all SNe Ia are produced via the SD scenario
  assuming $\dot{M}=10^{-7}M_{\odot}/\mathrm{yr}$, and an initial WD mass of $1.2 M_{\odot}$
  and the delay time distribution given by \citet{tmod+08}.
  The red dotted line is similar to the cyan dashed lines but assuming 
  $\dot{M}=3\times10^{-7} M_{\odot}/\mathrm{yr}$, an initial WD mass of $1.0 M_{\odot}$ (cf. \citealt{gb10}). 
  }
\label{fig:lumx}
\end{figure}

\begin{table*}
\caption{Comparison of the predicted X-ray luminosity in the soft X-ray band
  (0.3-0.7keV) in our standard model with observations. The observed X-ray luminosities for
  individual galaxies are taken from \citet{bg10,zgb12}.}
\label{tab:nsl}
\begin{tabular}{lccccccc}
\hline 
Name  & age  & $L_{\rm K}$  & $N_{\rm H}$ & $L_{\mathrm x}$ observed & $L_{\mathrm x}$ predicted& $L_{\mathrm{x,predict}}/L_{\mathrm{x,observed}}$  & SSSs Num\\
      & (Gyr)& ($L_{\rm K,\odot}$) & ($\times10^{20}\rm{cm}^{-2}$) & (erg/s) & (erg/s) & \\
(1)   & (2)  & (3)  & (4) & (5) & (6) & (7) & (8)\\
\hline 
NGC3585 & 3  & $1.5\times10^{11}$  & 5.6 &$3.8\times 10^{38}$ & $(2.6-3.0)\times 10^{39}$& $6.8-7.9$ & 330-570\\ 
\hline 
M32   & 4-10 & $8.5\times10^{8}$ & 6.3 &$1.5\times 10^{36}$ &$(1.2-1.7)\times 10^{36}$ & $0.8-1.1$ & 0 \\ 
\hline 
NGC3377 & 4  & $2.0\times10^{10}$ & 2.9 &$4.7\times 10^{37}$ &$(2.0-2.2)\times 10^{38}$ & $4.3-4.7$ &30-60 \\ 
\hline 
M31 bulge& 6-10 & $3.7\times10^{10}$  & 6.7 & $6.3\times 10^{37}$ &$(3.9-6.1)\times 10^{37}$& $0.6-1.0$ & 10 \\ 
\hline 
M105& 7-9 & $4.1\times10^{10}$ & 2.8 &$8.3\times 10^{37}$ & $(1.7-2.8)\times 10^{38}$& $2.0-3.4$ & 10-40\\ 
\hline 
NGC4278& 11 & $5.5\times10^{10}$  &1.8 &$1.5\times 10^{38}$ & $(0.88-1.98)\times 10^{38}$& $0.6-1.3$ & 10-40\\ 
\hline
\end{tabular}
\begin{tablenotes}
\small
\item {\bf Notes.} (1)-Galaxy name. (2)-Approximate stellar age of the galaxies. References-\citet{tf02,cmc09,obsd+06,sgcg06} (3) K-band luminosity of galaxies. 
(4)-Galactic absorption column density. (5)-Observed soft X-ray luminosity.
(6)-Predicted soft X-ray luminosity. (7)-Predicted soft X-ray luminosity over observed soft X-ray luminosity.
(8)-Predicted SSS number.
\end{tablenotes}
\end{table*}

Figure \ref{fig:lumx} shows the evolution of the mass-normalized X-ray
luminosity of the population of accreting WDs in the starburst and
constant SFR cases. For our predicted observable X-ray
luminosity, we apply absorption with a column density $N_{\rm H} =
3.0\times10^{20}\rm{cm}^{-2}$, which is a typical value for the
extinction within the Galaxy at high Galactic latitudes
\citep{dl90}. In addition, for early-type
galaxies we also show the observed X-ray luminosity per unit mass with
column density $1.8\times10^{20}\rm{cm}^{-2}<N_{\rm
  H}<6.7\times10^{20}\rm{cm}^{-2}$(the shaded region in the plot).
This range of $N_{\rm H}$ corresponds to that observed for the six
elliptical galaxies in Table \ref{tab:nsl}.  In the calculation, we
only considered Galactic absorption, since the intrinsic
absorption of these elliptical galaxies is small. 

Our calculations predict that in the starburst case, the X-ray
luminosity sharply decreases after 1 Gyr (Fig. \ref{fig:lumx}). At
early times (stellar age $t < 1.0$ Gyr), the X-ray luminosity in the
starburst case is comparable with that in the constant SFR case for
galaxies with the same mass. However, it is smaller for $t > 1.0$
Gyr. In order to compare our results with observations we use X-ray
data from \citet{bg10} and \citet{zgb12} for nearby elliptical
galaxies.  The data and our predictions are summarised in Table
\ref{tab:nsl} where we list the observed X-ray luminosities along with
the predicted X-ray luminosities and SSS numbers for six elliptical
galaxies \citep{bg10,zgb12}. Note that observed luminosities are total
soft X-ray luminosities of unresolved emission and soft compact
sources and therefore should be regarded as upper limits on the soft
X-ray luminosity of nuclear burning white dwarfs (see \citet{bg10} for
details).  In the calculation of the predicted X-ray luminosities and
SSS numbers, we have taken absorption into consideration for different
galaxies with different column density values.

One can see in Fig. \ref{fig:lumx} that, at the ages exceeding 6 Gyr,
our predicted X-ray luminosities are comparable to that which are
observed in nearby ellipticals, and is roughly two orders of magnitude
below the soft X-ray flux expected if all SNe Ia are produced via the
SD scenario (see \citet{gb10}).  This is consistent with our
prediction that the SD channel does not significantly contribute to
the SN Ia rate at late delay times (see Fig. 8 in Paper I).  However,
for the two youngest galaxies in the \citet{bg10} sample, NGC3585 and
NGC3377, our standard calculations predict up to an order of magnitude larger
soft X-ray luminosities than observed.  The strong dependence of the
X-ray luminosity on the age of the stellar population predicted by our
calculation seems to contradict observations. In interpreting this
result one should keep in mind that there is some controversy
regarding the ages of the two youngest galaxies. \citet{isd07} found
that the age of NGC3377 is around 7.8~Gyr and \citet{ggp12} listed
literature values of the age of NGC3377, varying from 3.5~Gyr to
8.9~Gyr. In the case of NGC3585, \citet{mich06} found a single stellar 
population age of 1.7~Gyr, while \citet{hzkg+07} suggested that its age should be larger
than 3~Gyr.  On the other hand, metallicity may be important to
explain the discrepancy between model predictions and observations.
\citet{op04} and \citet{ggp12} found that the metallicities of the two
youngest galaxies, NGC3585 and NGC3377, are likely to be 
lower than solar metallicity. However, solar abundance is adopted in our
calculation. The metallicity plays a complicated role in the evolution
of accreting WDs. The metallicity influences the initial-final mass
relation for stars, the stable burning limits of accreting WDs, and
the evolution of mass transfer rates
\citep{unyw99,mch08,dpy08,kskp+13}. Therefore, any strong conclusion
based on the example of these two galaxies may be premature.  On the
other hand, a similar discrepancy was also found in comparing
predicted populations of low mass X-ray binaries with observations,
based on a larger and different sample of galaxies and age
determinations \citep{zgb12,fltt13}.

\subsection{UV emission of accreting white dwarfs}
\label{sec:uvem}

Since nuclear-burning white dwarfs are characteristically high
temperature sources, the emission from this population will ionize the
ISM and will have an important influence on the structure of the
ionized gas and emission lines of a galaxy. In particular, compared
with the single stellar population, accreting WDs are likely to
emit predominantly beyond the He II photoionizing limit (see Fig.1 in
\citealt{wg13}). So in this section, we will investigate the H
ionizing and He II ionizing luminosity of accreting, nuclear-burning
WDs as a binary population. In addition, we will compute the predicted ratio
of two recombination lines produced in the ISM, He II
$\lambda$4686\AA/H$\beta$, and compare this with observations.

In our calculation, we assumed the binary fraction to be 50\%.
  It has been suggested that the binary fraction may depend on the
  binary parameters \citep{kbgp+09,kh09,sddl+12}. We neglect this
  effect, and therefore may slightly underestimate the binary fraction
  in our calculations. According to its standard definition, the
  binary fraction includes all binary stars, irrespective of
  their separation and evolutionary state. Our population synthesis calculations keep
  track of only those binaries in which one of the components is a
  mass accreting white dwarf. Such mass-transferring binaries are in
  fact a very small minority in the population (in terms of the number
  of systems and total mass). In accounting stars in wide binaries we
  assume that their emission is similar to the
  emission of single stars of the same spectral type. To compute
  emission of single stars and stars in wide binaries we use the
  stellar population synthesis code of \citet{bc03}, as described
  below. In computing the normalisation of the stellar emission we use
  the total mass of stars, ignoring the small fraction ($0.7\%$ by mass ) of stars in
  mass transferring close binaries. Hereafter, we refer to the stars in wide
binaries and single stars as single population (SP) and accreting,
nuclear burning WDs as binary population (BP).  The total emission of
the entire population is computed as a sum of emission of the SP and
BP. The SP-only model represents the hypothetical case of
absence of mass-accreting white dwarfs in the stellar population.

\begin{figure}                                      
\includegraphics[width=84mm]{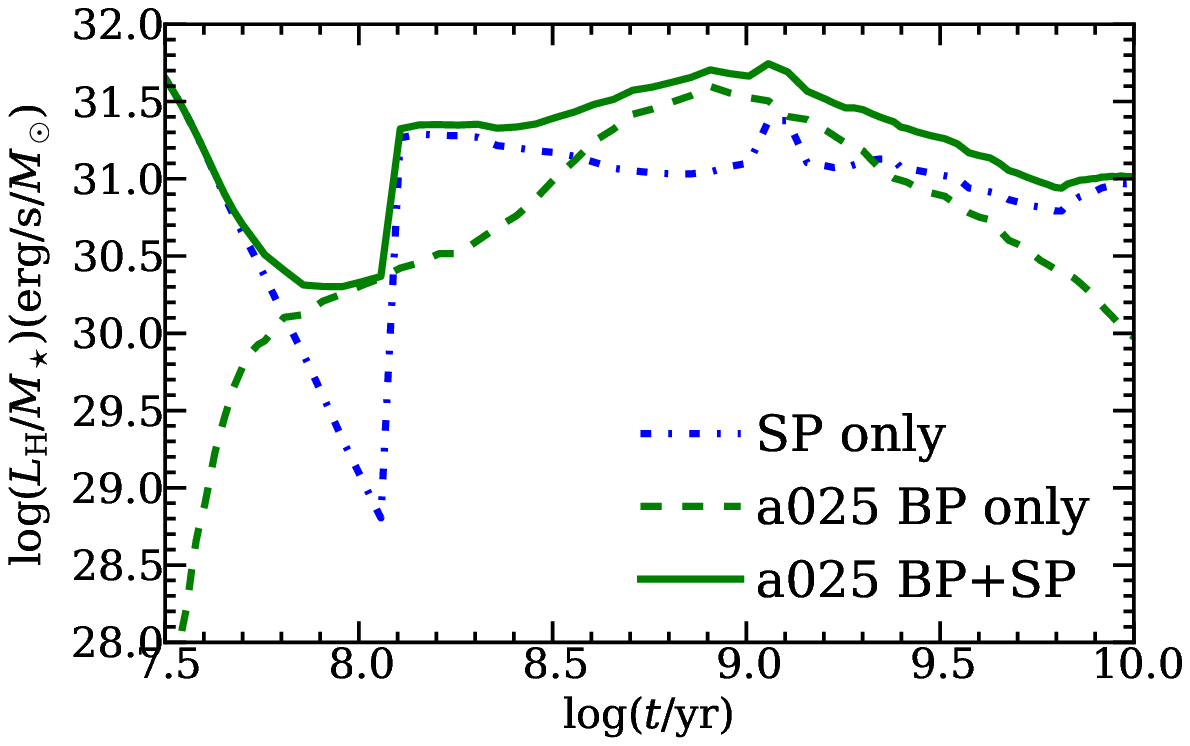}
\includegraphics[width=84mm]{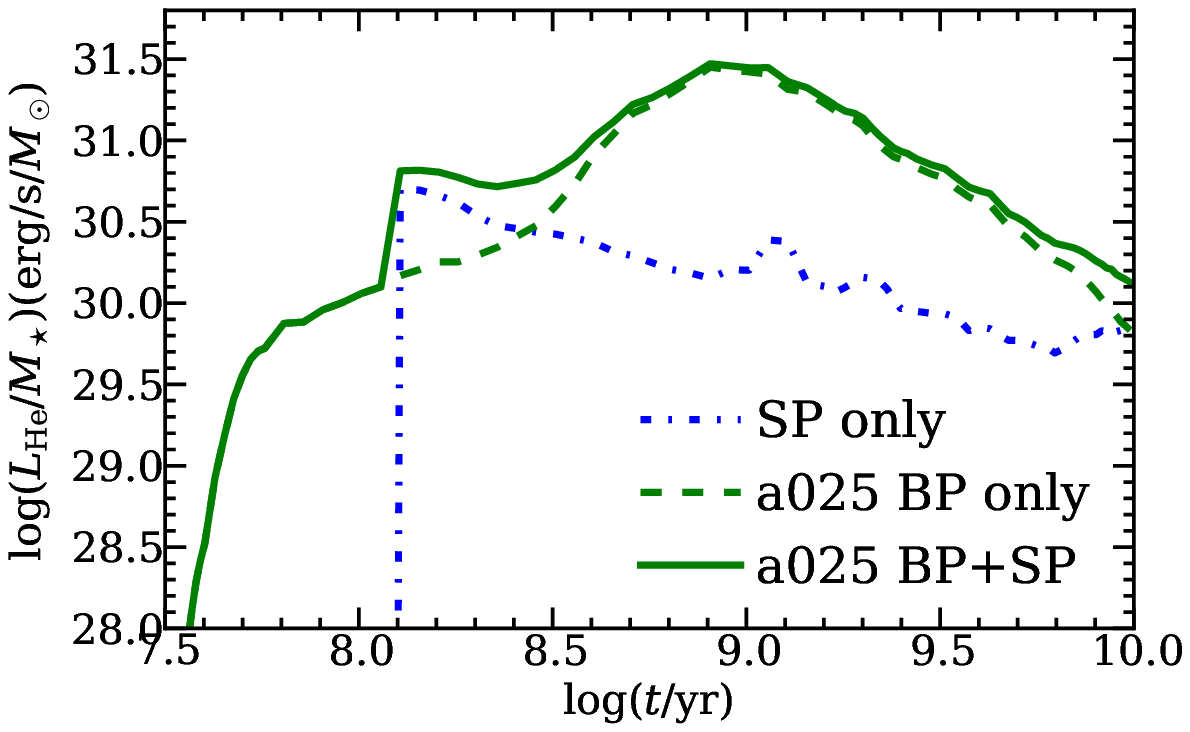}
\caption{Upper panel:H-ionizing ($h\nu > 13.6$ eV) luminosity per unit
  mass assuming ionization by single stars alone (SP only, blue
  dash-dotted line), ionizing radiation of a population of accreting
  white dwarfs for BP model a025 (a025 BP only, green dashed line),
  and their combined ionizing luminosity (green solid line) as a
  function of galaxy age in starburst case. Lower panel: similar to the upper
  panel, but for He-ionizing ($h\nu > 54.4$ eV) luminosity in starburst case. 
  }
\label{fig:lumhhe}
\end{figure}

In Fig. \ref{fig:lumhhe}, we show the evolution of the H-ionizing and
He-ionizing luminosities as a function of stellar age in the starburst
case. For the binary population, we only show the results for model
a025, since we do not find a significant difference between models
a025 and a050. To compute emission from the SP, we use the
  \citet{bc03} stellar population synthesis model. In running their
  model, we assumed solar metallicity and \citet{chab03} IMF, as this
  model is not available for Kroupa IMF.  This introduces only small
  inconsistency with our BP calculations (conducted for the Kroupa
  IMF).  The Chabrier IMF has the same exponent as the Kroupa IMF for
  stellar masses above 1.0 $M_{\odot}$ and a more complicated shape at
  smaller masses.  If the two IMFs are normalised to the same total
  mass, their high mass ($>1.0 M_{\odot}$) parts have the same shape
  and only small ($\sim 7\%$) difference in normalisation. This will
  result in a similarly small difference ($\sim 7\%$) in the predicted
  ionizing luminosity between the two IMFs, since stars with initial
  mass below 1.0 $M_{\odot}$ do not contribute to the ionising UV
  radiation. Such a small difference is insignificant given the large
  dynamical range of the ionising luminosities plotted in
  Fig. \ref{fig:lumhhe}.  In Fig. \ref{fig:lumhhe}, we do not show
the case of a constant SFR, as in this case the H-ionizing and
He-ionizing emission from accreting white dwarfs are significantly
smaller than emission from the SP. This is due to the fact that in the
constant SFR case there are many luminous young stars with high
effective temperatures, which dominate the ionizing emission.

In the starburst case, the H-ionizing emission from the population of
accreting WDs is comparable to that from the SP in the $\sim 0.3-2.0$
Gyr age range; it is unimportant for much earlier ages and for old
galaxies. The He-ionizing emission on the contrary is predicted to
dominate the ionizing UV background at ages greater than $\sim 0.3$
Gyr. In the SP model, massive stars dominate the ionizing UV
  radiation at early times. As the galaxy age increases, the main
  sequence turn-off mass decreases and the effective temperatures of
  the hottest stars in the population drops.  This leads to the
  decrease of the ionizing luminosity seen in the Fig.\ref{fig:lumhhe}
  at $t\la 100$ Myr.  Around $t\sim 100$ Myr, the first post-AGB stars
  appear, resulting in the sharp increase in the ionizing luminosity
  at $t\sim 100$ Myr, thereafter they dominate the UV emission. Note
  that in the SP, the He-ionizing luminosity provided by the massive
  stars drops very steeply and is outside the plotting range in the
  lower panel of Fig.~\ref{fig:lumhhe}. In the considered time range
  ($t\ga 30$ Myrs), the He-ionizing luminosity of the SP becomes
  significant only after the first post-AGB stars appear at $t\sim
  100$ Myr.  Note also, that the little ``bump'' around $1$ Gyr,
which can be seen on the curves in Fig.~\ref{fig:lumhhe}, is due to
the fact that $\sim2M_{\odot}$ stars undergo a helium flash and their
effective temperatures increase.

\citet{wg13,wg14} proposed that the UV emission from accreting WDs is 
capable of ionizing the ISM. They investigated the contribution of 
accreting, nuclear-burning WDs in the SD scenario (as SNe Ia progenitors) 
and pAGB stars to the ionizing background and
their influence on the emission lines of warm ISM in early-type galaxies.
They predicted that, if the SD scenario for SNe Ia is the dominant channel, 
SNe Ia progenitors should strongly ionize He, producing strong
He recombination lines in the extended emission-line regions of early-type galaxies.
Such lines are predicted to be much weaker if only single pAGB
stars are considered. Using a similar line of reasoning, one implication of
the large population of accreting WDs found in our results across a large range of delay-times
is that there should be strong He II recombination lines (e.g. He II 4686$\AA$) 
detectable in stellar populations with ages
larger than 0.3 Gyr but less than $\sim 5$ Gyr.

Among the recombination lines of He II, He~II~$\lambda$4686 is the
strongest line in the optical band, and is not heavily absorbed
by the ISM. In order to avoid the uncertainties in the covering
factor, we directly compare the predicted ratio of He~II~$4686
\AA/\rm{H}\beta$ with observations. In order to predict the emission
line fluxes, we make use of the photoionization code MAPPINGS III
\citep[e.g.][]{kdsh+01,gvs04}. The procedure is similar to that in
\citet{wg13}. Here we briefly summarize the main points and
assumptions in the calculation.

\citet{soma+12} found that the common morphology of HI gas in
elliptical galaxies is a regular HI disc or ring. The HI discs may
extend to several kpc well beyond the nucleus, and may or may not be confined within
the stellar body of the galaxy. Given that this gas will be ionized by
the combined diffuse emission of many distant sources, we assume plane-parallel
geometry in our calculation.  We adopt a constant hydrogen density
$n_{\rm H} = 100~\rm{cm}^{-3}$, which agrees with the observed ratio
of [SII] 6716\AA/6731\AA\, in passively-evolving galaxies
\citep{yb12}. Regarding the metallicities of the gas, which can be
estimated from the oxygen abundance, \citet{ab09} found an average
value of $Z_{\rm oxygen} \simeq Z_{\rm oxygen,\odot}$. So we use solar
abundances in our calculation.  Although the hydrogen density may be
varied as the galaxy evolves or the metallicity may be different, the
line ratio calculated in this work is relatively insensitive to the
density and the metallicity. Following \citet{jwgs+14}, we assume that
the ionization parameter $\rm{log}(U)\approx-3.5$, where
$U=\frac{\dot{N}_{\rm ph}}{4\pi r^{2}n_{\rm H}c}.$

\noindent This is consistent with the [O III] $\lambda 5007$/H$\beta$
ratio found in their stacks of early-type galaxies.

\begin{figure}                                      
\includegraphics[width=84mm]{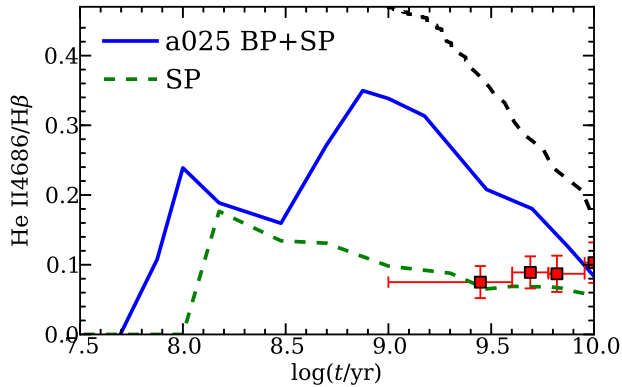}
\caption{Comparison between the predicted values of He II
  $\lambda4686/\rm{H}\beta$ in our starburst models and that which is
  observed in (stacks of) early-type galaxies, as a function of
  stellar age. The blue solid line shows the predicted values of He II
  $\lambda 4686/\rm{H}\beta$ for the combined populations (a025
  BP+SP) and the green dashed line is for the single stellar
  population from \citep{bc03}.  The black dashed line shows the
  predicted values of He II $\lambda 4686/\rm{H}\beta$ for the model
  combining SNe Ia progenitors in SD-scenario and post-AGB stars
  (similar to the model in \citet{wg13}).  In the calculation of the
  emission of SNe Ia progenitors, we assumed that all the SNe Ia 
  are produced via SD scenario, the initial WD mass $1.1
  M_{\odot}$, WD effective temperature $T_{\rm eff}=2\times10^{5}$K
  and the delay time distribution given by \citet{tmod+08}.  The
  observed values (red squares) are data from \citet{jwgs+14}.  Note
  that for these points the vertical bars denote the error in the
  observed value, but the horizontal bars simply indicate the width of
  each age bin.  }
\label{fig:HeIIHb}
\end{figure}

In Fig.~\ref{fig:HeIIHb}, we present the computed 
He~II~4686\AA/H$\beta$ ratio powered by the emission from the SP together
with the accreting WD population, and powered by the emission of the
SP alone. At early times, the ratio begins to increase with the
decrease in H-ionizing luminosity and increase of the He-ionizing
luminosity. The ratio reaches its first maximum when the H-ionizing
luminosity reaches its minimum.  Around 0.15~Gyr, the H-ionizing
luminosity from the SP increases and exceeds the emission from the
accreting WDs by an order of magnitude, leading to the beginning of a
decrease in this ratio.  After that, the H-ionizing luminosity does
not change significantly. Thereafter, the evolution of this ratio
simply reflects the evolution of the He ionizing luminosity.  The
luminosity ratio at 10 Gyr is consistent with observations, but the
computed ratio is larger than the observed one by a factor up to
$\simeq 3$ in the age range $1-8$ Gyr. This comes about due to the
predominance of accreting WDs in producing our predicted ionizing
background at these delay times. Although the total luminosities of
accreting WDs and post-AGB stars in this age range are comparable, the
lack of a sharp cutoff at 54.4eV in the spectra of accreting WDs leads
to a dramatically greater production of He II-ionizing photons. In
particular, the He II-ionizing luminosity predicted in our a025 model
exceeds that in the SP case by up to an order of magnitude at t$\sim$
1Gyr (see Fig. \ref{fig:lumhhe}).

\subsection{Emission from subsets of accreting WDs}
\label{sec:esub}

\begin{figure}                                             
\includegraphics[width=84mm]{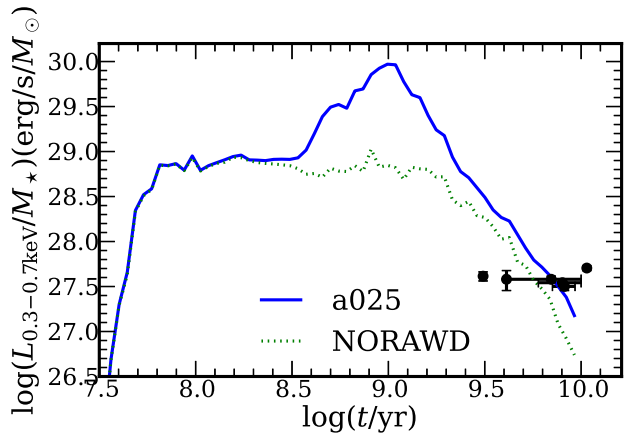}
\caption{Dependence of the soft X-ray luminosity (absorption applied
  with $N_{\rm H} = 3.0\times 10^{20}\rm{cm}^{-2}$) from different
  models of accreting WDs on the stellar ages in starburst case. The
  blue solid line is for the model a025. The green dotted line is for
  the model NORAWD assuming that accreting WD will enter CE instead of
  RAWD phases.}
\label{fig:lumx_diff_type}
\end{figure}

\begin{figure}                                           
\includegraphics[width=84mm]{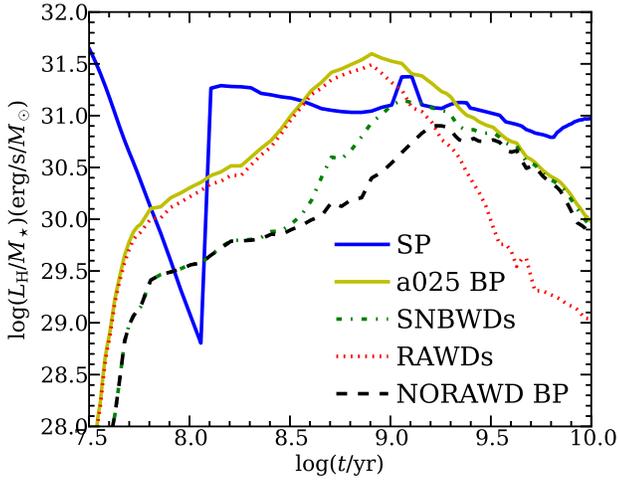}
\includegraphics[width=84mm]{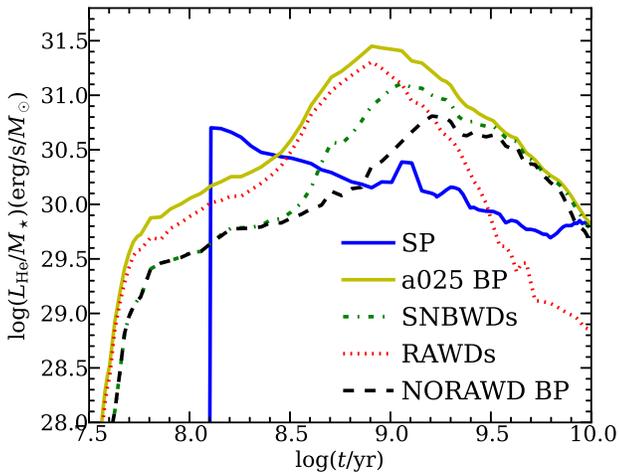}
\caption{Upper panel: Evolution of H-ionizing luminosity from different
  subsets of accreting WDs as a function of stellar age in the starburst case. 
  The blue solid line is for the SP model. The yellow solid line is for model
  a025. The green dash-dotted line and red dotted line are for the
  contribution of the SNBWDs and RAWDs, respectively. 
 The black dashed line is for the model that assumes the accreting WDs
  will enter a CE instead of a RAWD phase.  Lower panel: Similar
  to the upper panel, but for He-ionizing luminosity.}
\label{fig:lumhhe_diff_type}
\end{figure}

\begin{figure}                   
\includegraphics[width=84mm]{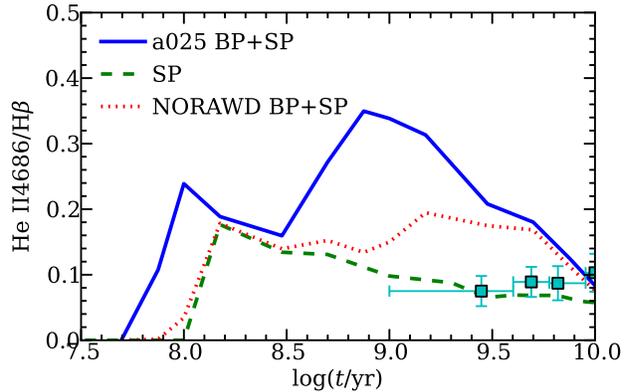}
\caption{The evolution of the HeII/H$\beta$ ratio as a function of
  stellar age for different model assumptions in the starburst case. 
  The blue solid line and red dotted 
  line show the values for the combined population in model a025 and
  NORAWD, respectively. The green dashed line and square data are
  the same as Fig. \ref{fig:HeIIHb}.  
 }
\label{fig:HeHb_worawd}
\end{figure}

In Fig. \ref{fig:lumx_diff_type}, we show the soft X-ray emission from
SNBWDs in model a025 and in an additional model in which we assume
that upon exceeding the maximum accretion rate for stable burning,
accreting WDs enter a CE phase instead of becoming RAWDs (model
NORAWD) and WDs cease to contribute to X-ray and UV output of the
population.  Similar to Fig. \ref{fig:lumx_diff_type}, we show the
H-ionizing luminosity and He-ionizing luminosity in
Fig. \ref{fig:lumhhe_diff_type}.  For the young stellar populations
with age $t < 1$Gyr, the RAWDs dominate the H and He II ionizing
emission.  For old stellar populations ($t > 1$ Gyr), the H and He
ionizing luminosity is dominated by emission from the SNBWDs. This is
due to the fact that, at early times, the binaries have massive donor
stars and higher thermal timescale mass transfer rates.  In contrast,
at later times, the donors are less massive and the thermal mass
transfer rate is lower, leading to more emission from the SNBWDs.  In
the model NORAWD, the soft X-ray luminosity, H-ionizing luminosity and
He-ionizing luminosity are very different from that in model a025. In
Paper I (see Fig. 2), we found that there are two stable burning
phases for some accreting WD binaries. In model NORAWD, replacing the
RAWD phase with a CE, only the first stably burning phase will
contribute to the emission.  So the H and He-ionizing luminosity in
model NORAWD are comparable to the emission from SNBWDs at early and
late times, when massive systems do not make a contribution.  In
Fig. \ref{fig:HeHb_worawd}, we show the computed HeII
4686\AA/$\rm{H}\beta$ ratio in the model NORAWD and in model a025.
The ratio reduces significantly, but it is still larger than the
observed ones by a factor of 2.  This corresponds to a remaining
discrepancy of up to factors of 2 to 5 between our prediction for the
He II-ionizing luminosity and that expected from the single stellar
population alone (see Fig.~\ref{fig:lumhhe}).


\section{Discussion}
\label{sec:disc}

\subsection{Model uncertainties}
\label{subsec:unc}

In the section above, we found that there is a strong dependence of soft
X-ray luminosity $L_{\rm 0.3-0.7keV}$ and the ratio He II $4686
\AA/\rm{H}\beta$ on stellar age for non-star-forming early-type galaxies in our
calculation, in contrast with observations (Fig. \ref{fig:lumx} and
Fig. \ref{fig:HeIIHb}). 
In producing a population synthesis model,
there exist several uncertainties in both the initial conditions and
the treatment of stellar evolution which may influence the results.
In the following, we discuss a (partial) list of several aspects which
may significantly impact our model, and their viability in providing
a solution. Some of the model parameters not considered here were
either tested and found to be of minimal importance (e.g. IMF,
mass ratio), or are simply unlikely to be of significant
impact (e.g. treatment of convective overshooting, as this only matters
for massive stars).

1) One important uncertainty is the retention efficiency of accreted
matter, (see \citet{btn13} \footnote{Note, in \citet{y10} He-retention
  efficiency is calculated differently from that quoted by
  \citet{btn13}.}.  In our calculation, we assumed that the He burning
retention efficiency is 100\%, which is clearly an overestimate and
results in an excess of massive WDs \footnote{ The WD mass may also be
  overestimated because of the initial-final mass function used in
  \textsc{BSE} code \citep[see][]{cpiv+14}.}, which may be partially
responsible for the high H and He ionizing emission of accreting
WDs. On the other hand, \citet{pty14} have shown that the conventional
assumption that the retention efficiency of matter at the surface of a
WD may be computed as a product of independently calculated retentions
of H and He (i.e. $\eta=\eta_{\rm H}\times\eta_{\rm He}$), results in
an underestimate of $\eta$. The reason is that H-burning heats the
underlying He layer and the outbursts of He-burning are milder than
that in the case of accretion of pure He.

\begin{figure}                                          
\includegraphics[width=84mm]{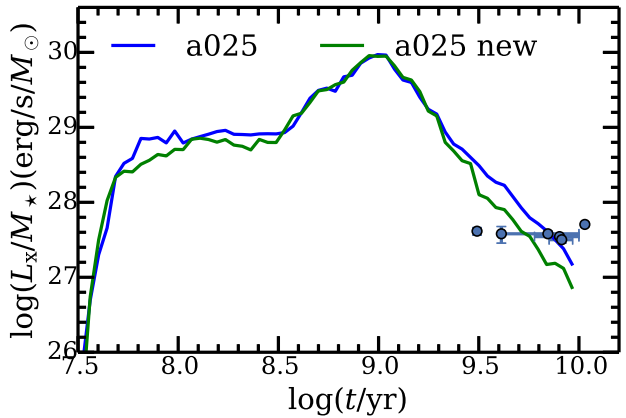}
\caption{Similar to the upper panel of Fig.~\ref{fig:lumx}. The ``new'' model a025  is 
the result computed using the revised He burning retention efficiency (see text).}
\label{fig:lumx_com}
\end{figure}

In order to check if He retention efficiency has a strong influence on our
result, we recomputed our results with the He burning retention efficiency given in 
\citet{hkn99}, and with the range in mass transfer rates corresponding to the 
stable burning regime as given in \citet{wbbp13}. In Fig.~\ref{fig:lumx_com},
we compare the soft X-ray luminosity from this new model with that from our standard model,
and find the difference is not significant. In spite of this, we should
emphasize that the retention efficiency suffers significant uncertainty, requiring further
investigation. Therefore, without more precisely understood retention efficiency,
we can not completely exclude the influence of this parameter.

2) In our calculations, solar abundance is adopted. As we discussed in
Section \ref{sec:xray}, some galaxies in our X-ray comparison are likely to
have low metallicities, which plays a complicated and still uncertain role in the evolution of
accreting WDs. On the one hand, at very low metallicities the stable-burning
regime may be extended to much lower mass transfer rates \citep{sb07}. However,
at the same time low-Z WD binaries are expected to have much higher accretion
rates \citep{ldwh00}, likely pushing many more 
binaries into dynamically unstable mass transfer. 
Either of these mechanisms require very low metallicities ($<$ 0.01$\rm{Z}_{\odot}$),
unlikely to be typical of many of the relatively nearby early-type galaxies used 
in our study. Therefore, we leave consideration of the evolution of WD binaries at 
low-Z to a future study. 

3) The X-ray comparison is based on a sample of six nearby galaxies
(the only currently available set of measurements suitable for such a
comparison). In addition, there is some controversy between different
age estimates for the two youngest galaxies. However, while the
incorrect age determination could in principle resolve the discrepancy
in the X-ray band, it is unlikely to explain the
He~II~4686~$\AA/\rm{H}{\beta}$ line ratio, as demonstrated in
\citet{jwgs+14}.

4) It may be that our modeling of the emission spectra of nuclear
burning white dwarfs is incorrect.  This must involve a failure of the
1D theory, rather than any simple deviations from the black body spectral
template used throughout the paper (see section \ref{sec:emi}). One
obvious possibility is that the photospheric radii may be much larger
than predicted by 1D models of the nuclear burning on the WD surface.
This would shift the peak of their emission to the UV or optical bands where
they would be remarkably unusual objects. However,
\citet{lv13} searched for such objects in the Small Magellanic Cloud
and found no plausible candidates. Therefore this possibility seems to
be unlikely.

5) One of the most important uncertainties in our calculations is the
treatment of the CE phase. There are still many aspects of CE
evolution which remain unclear, such as the available energy sources
and the definition of the core-envelope boundary (see \citet{ijcd+13}
for a review,\citet{ht14}). In addition, the criteria for the onset of a CE phase
also suffer from great uncertainty. In the following 
subsection, we will discuss this possibility in depth and provide 
a possible solution to resolve the discrepancy between our computed 
results and observations.

\subsection{A potential solution}
\label{subsec:psol}

In our calculations, for binaries 
harbouring giant donors, we adopted after \citet{hw87} and \citet{webb88}
as the criterion for stable mass loss 
a critical value of mass ratio (hereafter, HW criterion):

\begin{equation}
q_c = 0.362 + \frac{1.0}{3.0(1-m_c)},
\label{eq:mra}
\end{equation}
\noindent where $m_c$ is the core mass fraction. For conservative mass transfer,
this predicts dynamical instability for stars with a core mass fraction
$\lesssim 0.45$, i.e. $q_c \lesssim 1.0$. \citet{ch08} (see also \citet{hpmm+02}), using
detailed binary evolution calculations, found that the critical mass
ratio can be up to $\sim 2.0$, depending on the evolutionary phase of the
donor star, mass and angular momentum loss from the system.
 \citet{wi11} (see also
\citet{php12}) found that this difference is due to the existence of a
superadiabatic outer surface layer, which plays a critical role during
the mass transfer. Following on the work of \citet{wi11}, \citet{pi15}
found that the critical mass ratio varies from 1.5 to 2.2 for
conservative mass transfer. Given these recent results, we turn now to considering
alternatives to the HW prescription, in particular a more stringent 
condition for
the onset of a CE, as a possible solution to the overproduction of accreting WDs
found in our standard model above. We attempt to replace the HW
criteria with $q_c = 1.5$ (model a025qc15), $1.7$ (model a025qc17), and $1.9$ (model a025qc19),
 for binaries consisting of MS and giant stars. Although the mass ratio needed for the onset
of a CE is understood to vary significantly for differing combinations of accretor mass, 
donor mass, and evolutionary state of the donor, these values are more consistent with the results 
of detailed numerical simulations \citep[e.g.][]{pi15}, and much higher than what is typically
found when implementing the HW criterion.
In this section, using these new criteria, we calculated the soft X-ray, H-ionizing and
He-ionizing luminosity as a function of stellar age.

\begin{figure}                  
\includegraphics[width=84mm]{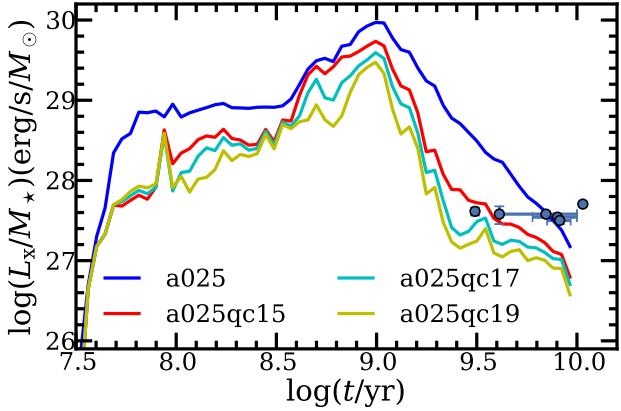}
\caption{Evolution of soft X-ray luminosity (absorption applied with
  $N_{\rm H}= 3.0\times10^{20} {\rm cm}^{-2}$) as a function of
  stellar age in starburst case. The red, green and yellow solid lines
  are for model a025qc15, model a025qc17 and model a025qc19,
  respectively. 
}
\label{fig:lumx_sol}
\end{figure}

\begin{figure}                  
\includegraphics[width=84mm]{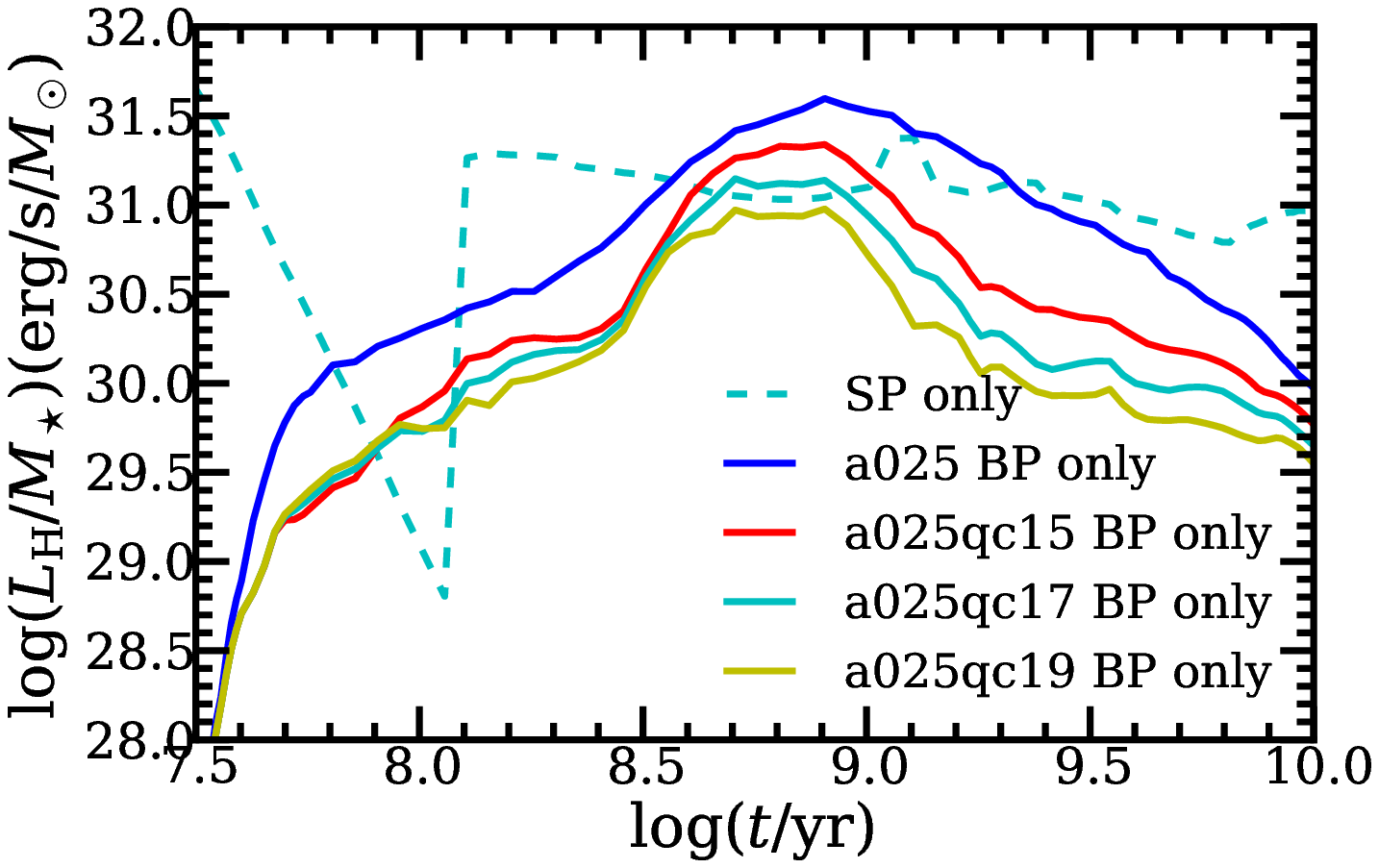}
\includegraphics[width=84mm]{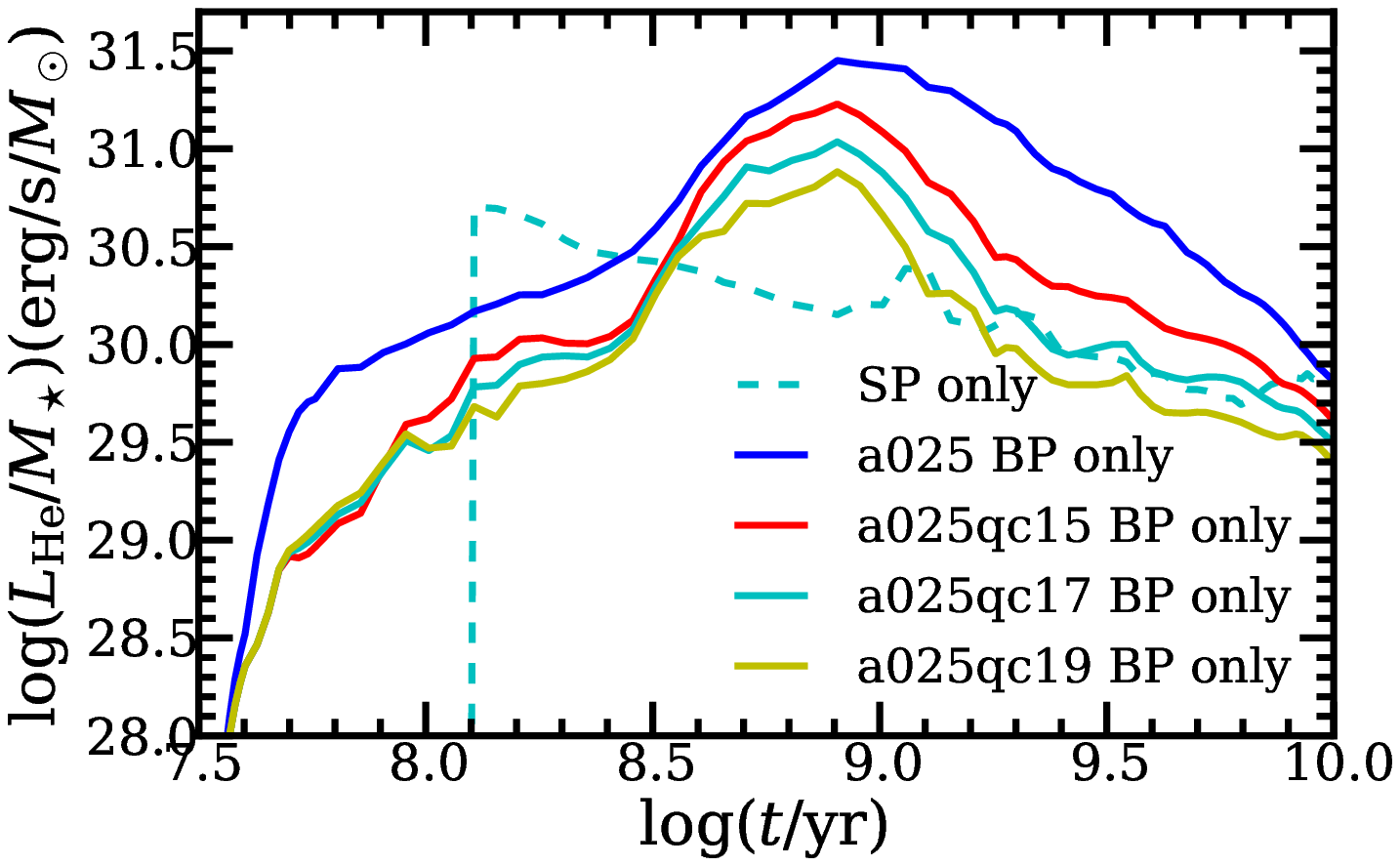}
\caption{Evolution of H-ionizing (upper panel) and He-ionizing (lower panel) 
  luminosity as a function of stellar age in different models. }
\label{fig:lumh_sol}
\label{fig:lumhe_sol}
\end{figure}

\begin{figure}                   
\includegraphics[width=84mm]{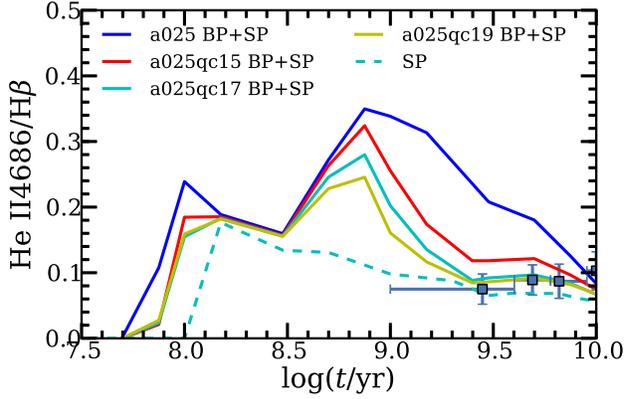}
\caption{Evolution of line ratio HeII 4686/H$\beta$ as a function of
  stellar age for combined population in model a025qc15, model a025qc17 and model a025qc19. }
\label{fig:ratio_sol}
\end{figure}

In Fig.~\ref{fig:lumx_sol}, we show the evolution of soft X-ray
luminosity as a function of stellar age. Compared with our model using 
the HW criterion, the new criteria have larger critical mass ratios.  With larger critical
mass ratios, the predicted soft X-ray luminosity becomes lower. This is due to
the fact that, with the requirement of a larger critical mass ratio, it will be relatively difficult for binaries
to enter dynamically unstable mass transfer.  Therefore, fewer binaries
will enter a CE phase, and fewer accreting WDs will form. For old stellar
populations, the predicted soft X-ray luminosities in model a025qc17 and model
a025qc19 are lower than the observed values.  Given that the
observational soft X-ray luminosity is the upper limit for the
emission of accreting WDs, our results are consistent with observations.  
Fig.~\ref{fig:lumh_sol} shows the evolution of the H-ionizing and He-ionizing luminosity for 
different models, respectively. These results are similar to model a025, but the
H-ionizing and He-ionizing luminosity is lower, for the same reason as above. In
Fig.~\ref{fig:ratio_sol}, we present the dependence of line ratio HeII
$\lambda 4686/\rm{H}\beta$ on stellar age for the new models. For
old stellar populations in model a025qc17 and model a025qc19, the line
ratio is much smaller than that in model a025 and consistent with
observations.

The use of fixed unique critical mass ratios for the onset of 
rapid, unstable mass transfer is not realistic \citep{nvyp00,wi11}.
Numerical experiments above show that stability criteria of mass loss 
for giant stars play a crucial role in defining the number of SNBWD. 
However, stability also depends on mass of the donor, 
specific momentum of matter lost from the system etc.
Fine-tuning of all these parameters may bring even better agreement of
model and observations, but it is computer-time prohibitive.

\subsection{Number of SSSs}
\label{subsec:nsss}

\begin{figure}                         
\centering
\includegraphics[width=84mm]{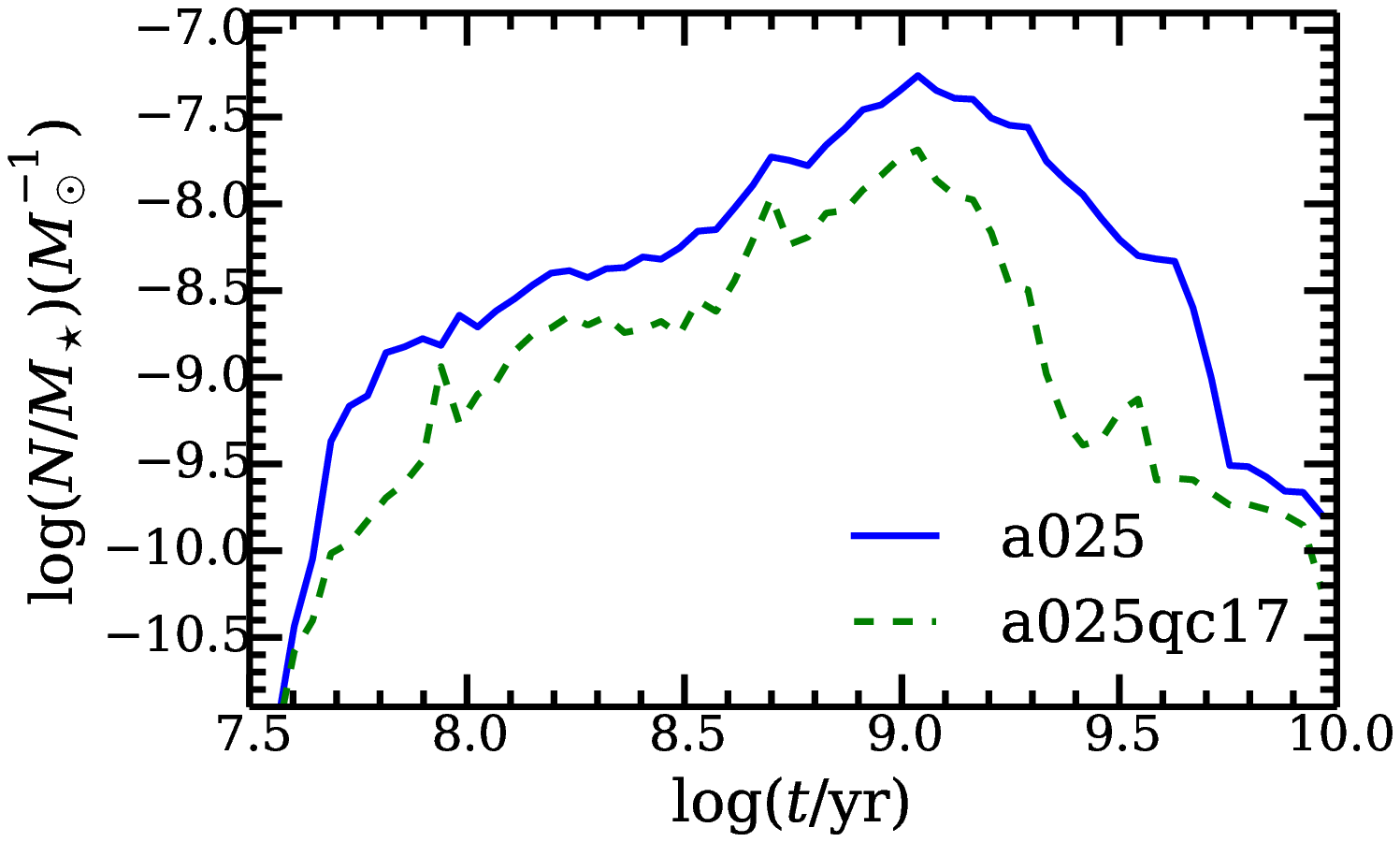}
\includegraphics[width=84mm]{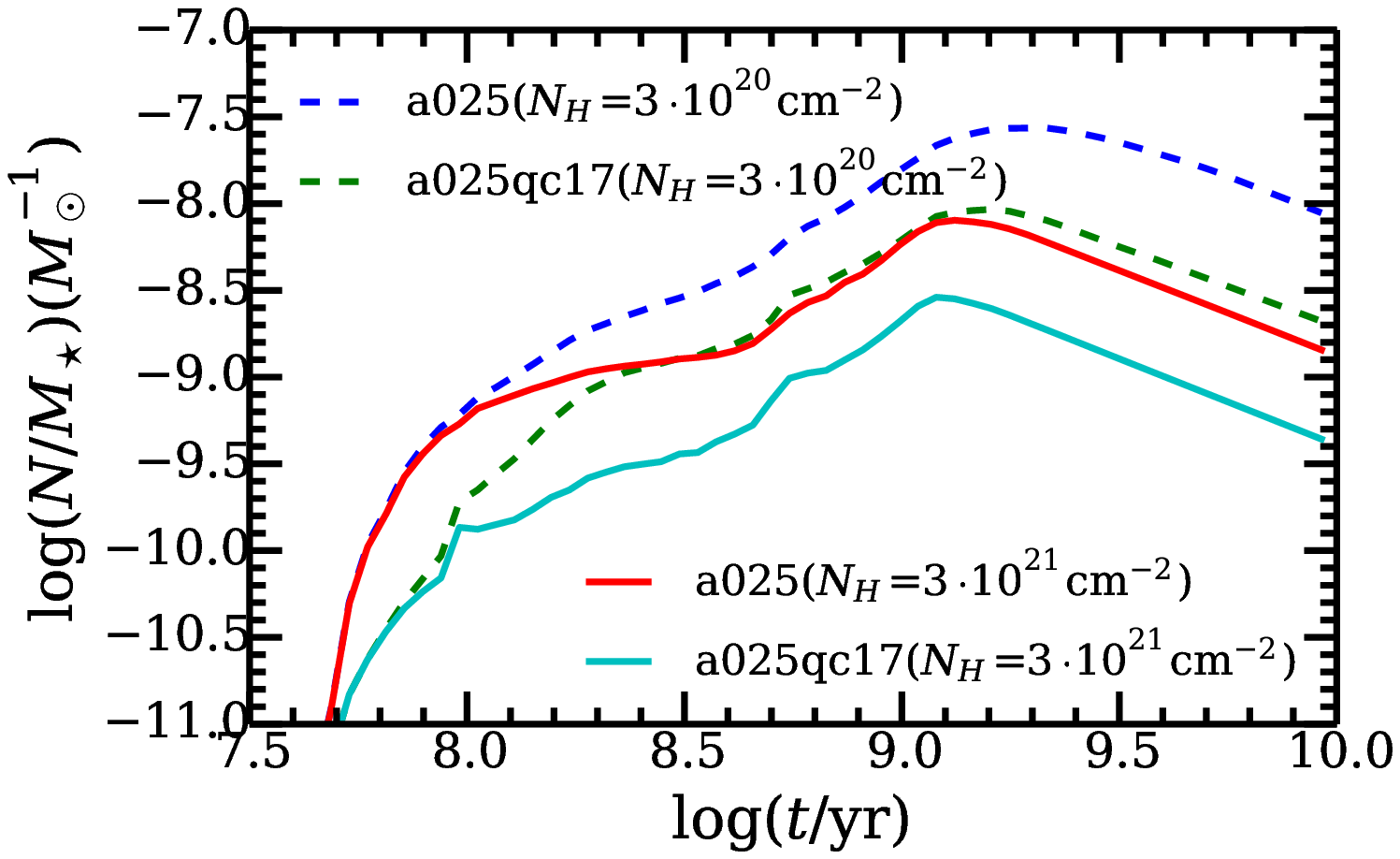}
\caption{Evolution of the number of SSSs per unit stellar mass in
  the starburst case (upper panel) and the constant SFR case (lower
  panel). The blue dashed line is for model a025 and green dashed line
  for model a025qc17 assuming $N_{\rm H}=3.0\times10^{20}\rm cm^{-2}$.      
  The red and cyan solid lines in the bottom panel show 
  the number of SSSs with $N_{\rm H}=3.0\times10^{21}\rm cm^{-2}$ in
   model a025 and model a025qc17. }
\label{fig:numsss}
\end{figure}

As discussed in Paper I, not all SNBWDs can be detected as SSSs, since
this is dependent on the luminosity and effective temperature of the
source, and the column density of the gas along the line of sight. In
this paper, we define SSSs as SNBWDs with soft X-ray (0.30-0.70keV)
luminosity $L_{\rm x}>10^{36}$erg/s. With this definition, we
computed the evolution of the number of SSSs, which is shown in
Fig.~\ref{fig:numsss}. In the starburst case, the SSS population peaks 
around $\sim 1$~Gyr and then declines by $\sim 2$ orders of magnitude
by the age of 10~Gyr. We note that the ionizing radiation from SNBWDs
has a peak at the same age range (see the figures above). A similar
peak was found in earlier studies of SNe Ia rates in the single
degenerate scenario \citep[e.g.][]{crb96,hp04,rbf09,y10}. The reason for this
is that main-sequence stars with mass close to $\sim 2~M_{\odot}$ have 
lifetimes $\sim 1$~Gyr and radii and luminosities that enable a relatively
long stage of thermal time-scale mass loss with the rate $\sim 10^{-6}
M_{\odot}$/yr, corresponding to the stable hydrogen burning regime (see 
Figs. 2 and 10 in Paper I)\footnote{If mass loss is artificially fine-tuned,
e.g., by the highly uncertain effect of mass-stripping effect of donors by an accretor wind 
\citep{hkn99}, the peak may be shifted to younger ages, e.g., \citet{mvdd10}.
The peak is smeared by the scatter in the actual parameters of WD+MS pairs -
masses, mass ratios of components, and initial separations.}.
 The number of SSSs for a $10^{11}M_{\odot}$
galaxy at 10 Gyr is $10-20$ in the starburst case and
$200-820$ in the constant SFR case assuming $N_{\rm H} = 3\times10^{20}
\rm cm^{-2}$. Given the high intrinsic absorption of spiral galaxies, 
we also made a set of calculations assuming 
$N_{\rm H} = 3\times10^{21} \rm cm^{-2}$. In this case, 
the SSS number is $40-130$ for a spiral-like
galaxy. Note, however, that the
observable number of SSSs depends on the absorption column density,
which in turn depends on the relative distribution of gas and SSSs in
any given galaxy. 

\subsection{Gaseous nebulae around SSSs}

Given the hard emission of SSSs, it is suggested that they should be
accompanied by ionized nebulae.  \citet{rckm94} modeled the
ionization and temperature structure of such nebulae and predicted
strong [OIII] $\lambda$5007 and He II $\lambda$4686 emission
lines. \citet{rrm95} searched for gaseous nebulae surrounding SSSs in
the Large and Small Magellanic Clouds. They found only one SSS,
Cal~83 \citep[first identified as a high-excitation nebula
  by][]{pm89}, with a detected nebula, with null detections among the
9 other known Magellanic SSSs. This means that either the
time-averaged luminosity and/or temperature of these sources must be
much lower than presently observed, or that the local ISM density
around SSSs are typically too low. In fact, given that typical ISM densities
are 1 -- 2 orders of magnitude below that observed for the Cal 83
nebula, the latter appears more likely (see Woods \& Gilfanov (submitted), 
who found a 70\% probability that $\lesssim 1$ SSS in the LMC would
have a detectable nebula). The situation is complicated further in the accretion-wind
scenario -- for SSSs in star-forming galaxies, a wind-blown bubble
will excavate a cavity in the ISM surrounding the source which is 10
-- 40 pc in radius \citep{bhbl07}. Consequently, the ionization
parameter and morphology of these nebulae will be much different.

Since the common morphology of HI gas in elliptical galaxies is a
regular HI disk or ring, individual sources are not embedded in the
ISM they ionize. With the above explanation and HI gas morphology in mind,
we would expect that the individual emission line luminosity in
elliptical galaxies should not be influenced by any complications
introduced by the impact of the accreting WD on any surrounding ISM,
as is the case in star-forming galaxies.


\section{Summary and Conclusions}
\label{sec:con}

Based on the results of population synthesis of accreting WDs from
Paper I and using simple assumptions regarding the emission of
accreting WDs, we study the number of observed SSSs, the soft X-ray
luminosity of accreting WDs, and their H and He~II ionizing
luminosity.  The main conclusions are as follows.

\begin{enumerate}

 \item 
 We compare predictions of our model with observations of nearby
elliptical galaxies in optical and X-ray bands. To this end we use the
He~II~4686~$\AA/\rm{H}{\beta}$ line ratio to characterise the hardness
and strength of the ionizing UV continuum.  In our standard model, using 
commonly (although not universally) adopted assumptions, our predictions
for the soft X-ray luminosity of old stellar populations are consistent with {\it Chandra}
observations of several nearby elliptical galaxies.  Likewise, for
stellar ages of $\sim 10$~Gyr, the He~II~4686~$\AA/\rm{H}{\beta}$ line
ratio from warm ISM predicted by our model is consistent with that
measured in the stacked SDSS spectra of retired galaxies.  However,
for stellar ages of $\sim 4-8$ Gyrs our model significantly
over-predicts the soft X-ray luminosity and the
He~II~4686~$\AA/\rm{H}{\beta}$ line ratio.

 \item
We discuss various possibilities to explain this controversy and
tentatively conclude that the most likely reason 
is that the classical \citet{hw87} criterion typically used for 
evaluating the onset of dynamically unstable 
mass transfer for giant stars predicts too low critical mass ratio. Replacement of
this criterion by fixed critical mass ratios ($q_c = 1.7$ or 1.9), brings the soft X-ray luminosity and line
ratio He II$\lambda 4686/\rm{H}\beta$ in the new models to
consistency with the observations. However, we stress that the critical mass
ratio needed for the onset of a common envelope is actually not a fixed value,
and varies in an as yet unspecified way as a function of other stellar parameters.
Further  effort is needed in this regard, in order to better reconcile the predictions of binary
population synthesis with observations.

 \item
In the starburst case, the population of SNBWDs, their combined soft
X-ray and UV output and the number of SSSs (SNBWDs with $L_{\rm x}
>10^{36}$erg/s) peak at $\sim 1$ Gyr and then strongly decline with
age, by $\sim (1-3)$ orders of magnitude.
Assuming $N_{\rm H}=3\cdot 10^{20}$ cm$^{-2}$, the number of SSSs present at
10 Gyr in the starburst case  is $10-20$ for 
a galaxy with mass $10^{11}M_{\odot}$. In the constant SFR case,
assuming $N_{\rm H}=3\cdot 10^{21}$ cm$^{-2}$, the SSSs number at 10 Gyr
is around $40-130$ for the same mass galaxy.

\end{enumerate}

\section*{Acknowledgments}

We would like to thank Bill Wolf for kindly providing us model data
for stable-burning WDs.  We are grateful to the \textsc{MESA} council
for the \textsc{MESA} instrument papers and website.  The TheoSSA
service (http://dc.g-vo.org/theossa) used to retrieve theoretical
spectra for this paper was constructed as a part of the activities of
the German Astrophysical Virtual Observatory.  HLC gratefully
acknowledges support and hospitality from the MPG-CAS Joint Doctoral
Promotion Program (DPP) and the Max Planck Institute for Astrophysics
(MPA). This work was partially supported by the National Natural
Science Foundation of China (Grant No.11390374), Science and
Technology Innovation Talent Programme of Yunnan Province (Grant
No. 2013HA005). The work is also partially supported by the Presidium
of the Russian Academy of Sciences program P-41 and RFBR grants
No.~14-02-00604 and 15-02-04053. LRY gratefully acknowledges warm hospitality and
support from MPA-Garching. MG acknowledges hospitality
of the Kazan Federal University (KFU) and support by the Rus-
sian Government Program of Competitive Growth of KFU.
ZWH thanks the support by the strategic
Priority Research Program "The Emergence of Cosmological Structures"
of the Chinese Academy of Science, Grant No. XDB09010202.  HLC
acknowledges the computing time granted by the Yunnan Observatories
and provided on the facilities at the Yunnan Observatories
Supercomputing Platform.

\bibliographystyle{mn2e} 


\label{lastpage}

\end{document}